\documentclass[10pt,letterpaper,aps,pra,twocolumn,english,showpacs,superscriptaddress]{revtex4-1}

\usepackage[hidelinks]{hyperref}
\usepackage{amsmath}
\usepackage{mathtools}
\usepackage{amsfonts}
\usepackage{footnote}
\usepackage{todonotes}
\usepackage[caption=false]{subfig}
\usepackage{xcolor}
\usepackage{graphicx}
\usepackage[squaren]{SIunits}

\hypersetup{colorlinks=true, citecolor=blue, linkcolor=blue, urlcolor=	blue}

\makeatletter
\def\tagform@#1{\maketag@@@{\ignorespaces#1\unskip\@@italiccorr}}%
\let\orgtheequation\theequation%
\def\theequation{(\orgtheequation)}%
\makeatother

\newcommand{\abs}[1]{\lvert#1\rvert}

\newcommand{\ket}[1]{|#1\rangle}
\newcommand{\val}[2]{\ensuremath{#1~\mathrm{#2}}}
\newcommand{\V}[1]{\boldsymbol{\mathbf{#1}}}
\newcommand{\unitV}[1]{\V{\hat{#1}}}
\newcommand{\magV}[1]{\abs{\V{#1}}}
\newcommand{\g}{\ensuremath{\mathit{g}}}
\newcommand{\ugrtHz}{\ensuremath{\mathrm{\mu}\g\mathrm{/\sqrt{Hz}}}}
\newcommand{\uradsrtHz}{\ensuremath{\mathrm{\mu rad/s/\sqrt{Hz}}}}
\newcommand{\ngrtHz}{\ensuremath{\mathrm{n}\g\mathrm{/\sqrt{Hz}}}}
\newcommand{\nradsrtHz}{\ensuremath{\mathrm{nrad/s/\sqrt{Hz}}}}
\newcommand{\Rb}{\ensuremath{\mathrm{^{87}Rb}}}

\begin{document}

\title{Dual-axis, high data-rate atom interferometer via cold ensemble exchange}

\author{Akash V. Rakholia}
\affiliation{Sandia National Laboratories, Albuquerque, NM 87185, USA}
\affiliation{Center for Quantum Information and Control (CQuIC), Department of Physics and Astronomy, University of New Mexico, Albuquerque, New Mexico, 87131, USA} 
\author{Hayden J. McGuinness}
\affiliation{Sandia National Laboratories, Albuquerque, NM 87185, USA}
\author{Grant W. Biedermann}
\email[Corresponding author: ]{gbieder@sandia.gov}
\affiliation{Sandia National Laboratories, Albuquerque, NM 87185, USA}
\affiliation{Center for Quantum Information and Control (CQuIC), Department of Physics and Astronomy, University of New Mexico, Albuquerque, New Mexico, 87131, USA}

\begin{abstract}
We demonstrate a dual-axis accelerometer and gyroscope atom interferometer, which forms the building blocks of a six-axis inertial measurement unit. By recapturing the atoms after the interferometer sequence, we maintain a large atom number at high data-rates of 50 to 100 measurements per second. Two cold ensembles are formed in trap zones located a few centimeters apart, and are launched toward one-another. During their ballistic trajectory, they are interrogated with a stimulated Raman sequence, detected, and recaptured in the opposing trap zone. We achieve sensitivities at $\ugrtHz$ and $\uradsrtHz$ levels, making this a compelling prospect for expanding the use of atom interferometer inertial sensors beyond benign laboratory environments. 
\end{abstract}

\date{\today}
  
\maketitle
 
\section{Introduction}  
  
Light-pulse atoms interferometers (LPAIs) \cite{Riehle91,Kasevich91, Dickerson} have demonstrated remarkably precise and stable measurements of acceleration and rotation. Numerous applications have been suggested, including gravimetry \cite{pyramid}, seismology, inertial navigation \cite{McGuirk,Canuel06,Wu}, and near-surface force characterization \cite{Parazzoli}. However, such systems are typically designed for laboratory environments in pursuit of maximum sensitivity, and thus are bulky and fragile, operating at the few Hertz rate. Lately, there has been great interest in adapting such systems to field use \cite{Bidel13, Muller09}. Adapting LPAI technology to this application space requires a compact sensor explicitly tailored for dynamic environments. Furthermore, inertial navigation requires a sensor capable of measuring both accelerations and rotations. Building on our previously reported high data-rate accelerometer \cite{McGuinness}, we report a compact LPAI that is capable of simultaneous acceleration and rotation measurements, and suitable for dynamic environments. 

Our LPAI uses stimulated Raman transitions between hyperfine levels in a $\pi/2 - \pi - \pi/2$ pulse sequence \cite{Berman} to coherently separate, redirect, and recombine atomic wavepackets from a cold atomic sample. The spatially-dependent phase of the light field is imprinted on the atoms during each pulse. To first order, the resulting phase difference between the hyperfine levels is given by $\Delta\phi = \V{k_e} \cdot (\V{a} - 2\V{v}\times\V{\Omega})T^2$, where $\V{k_e}$ is the effective Raman wavevector, $T$ is the delay between pulses, and $\V{v}$, $\V{a}$, and $\V{\Omega}$ are the velocity, acceleration, and rotation of the atoms relative to the platform, respectively. Since $\V{k_e}$ is typically locked to an atomic transition, a precise determination of $\Delta\phi$ results in a precise measurement of acceleration and rotation. 

\begin{figure}
\centering{
	\includegraphics[width=1\columnwidth]{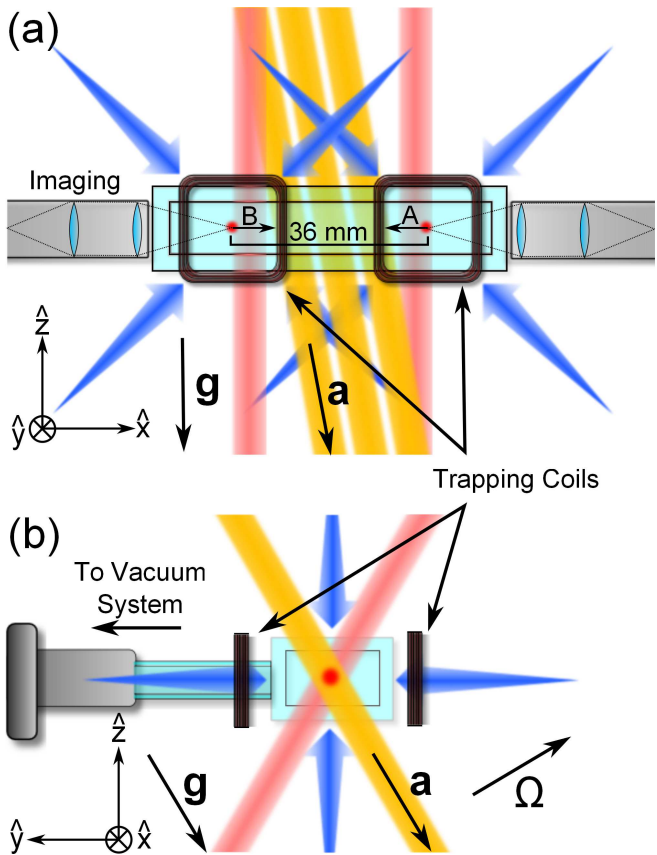}
	\caption{\label{fig:apparatus} (Color online) Diagram of the apparatus implementing the cold ensemble exchange and dual-axis, high data-rate atom interferometer. (a) Front view: Two MOTs are loaded \val{36}{mm} apart. Cooling beams are shown in blue, probe beams in pink, and Raman beams in yellow. The trap is turned off, and the outer and inner cooling beams are blue and red detuned, respectively, which launches the ensembles towards each other. After the experiment, atoms are recaptured in the opposite trap to facilitate loading. (b) Side view: The design allows for four planes of optical access, enabling a compact apparatus. The vector $\V{g}$ shows the direction of gravity, while $\V{a}$ and $\V{\Omega}$ are the directions of acceleration and rotation measurement, respectively.}
}
\end{figure}

\begin{figure}
\centering{
	\includegraphics[width=1\columnwidth]{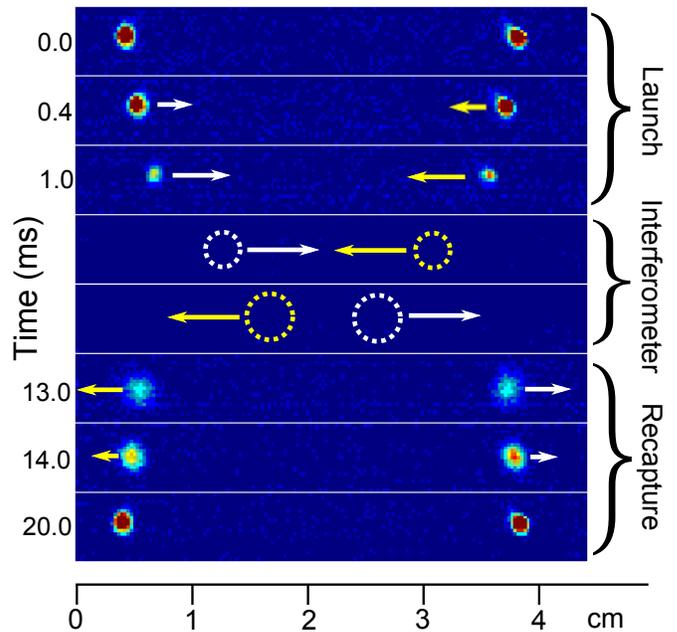}
	\caption{\label{fig:flight} (Color online) Frames from a movie demonstrating the ensemble exchange with a cycle time of \val{20}{ms}. The ensembles are launched towards each other over \val{2}{ms} to a velocity of \val{2.5}{m/s}, with simultaneous sub-Doppler cooling. The interferometer takes place during their ballistic trajectory with an interrogation time \val{T \approx 4}{ms}. Imaging beams are off during the interferometer to reduce photon scatter which would otherwise cause decoherence. Dashed circles represent the approximate positions of the ensembles during this stage. Atoms are then recaptured in the opposite trapping region, with additional atoms loaded from vapor for \val{7}{ms}. This cycle represents a high atom number, low bandwidth cycle to optimize the image contrast. Images were taken with a Lumenera CCD camera (Lm075) using a \val{0.2}{ms} exposure.}
}
\end{figure}

The phase shift $\Delta\phi$ includes both acceleration and rotation contributions. If we consider a system of two independent interferometers \cite{Gustavson00} with phase shifts $\phi_a$ and $\phi_b$, and opposing velocities $\V{v_a} = - \V{v_b}$, the sum and difference of the phases result in a term proportional to acceleration and rotation, respectively.

In a dynamic environment, a free-fall measurement such as this requires a short interrogation time \cite{Butts} to curtail otherwise significant and uncontrolled excursions of the ensemble. Due to the quadratic dependence of $\Delta \phi$ on $T$, a significant loss of sensitivity ensues. This can be compensated by using a large atom number and minimizing the dead time between measurements. We achieve both through recapture, recycling as many atoms as possible between shots \cite{McGuinness}. Given the constraint of a small $T$, it is preferable to maximize velocity to maximize the rotation rate sensitivity. One method to achieve a large velocity while maintaining a high data-rate is through the use of our ensemble exchange technique, reported herein.

\section{Apparatus}

Our ensemble exchange apparatus swaps cold atomic ensembles between two regions via launch and recapture. Two ensembles are simultaneously loaded in magneto-optical traps (MOTs) located a few centimeters apart. Through the use of conventional fountain launch techniques, the ensembles are launched towards each other at a velocity of a few meters per second. The interferometer takes place over several milliseconds during the ballistic transit between the two loading regions. At the end of the experiment, atoms are \emph{recaptured} in the opposite trap from where they were launched. After a few milliseconds of recapture/loading, the cycle repeats. Thus, the atoms are ``exchanged'' between the trapping regions. Typically, the background-limited time it takes to load a MOT from vapor is on the order of a second. With recapture, the loading time is determined by the trap's time-constant, $\tau_{\mathrm{MOT}}$, which is typically on the order of several milliseconds \cite{Metcalf}. 

We demonstrate this concept with a custom-built sensor head. The sensor head is designed around a rectangular quartz vacuum cell with inner dimensions of \val{20 \times 30 \times 60}{mm^3} and \val{3}{mm} thick walls (\autoref{fig:apparatus}). The use of high vacuum, as opposed to ultra-high vacuum, significantly relaxes the vacuum requirements for the apparatus. Two magneto-optical traps of $\Rb$ are loaded \val{36}{mm} apart at opposite ends of the cell from a background pressure of \val{2 \times 10^{-7}}{Torr}. The high vapor pressure is generated from a temperature-controlled source containing Rubidium from a crushed ampoule. The high background vapor density allows for a fast short-term loading rate of \val{1 \times 10^{5}}{atoms/ms}. This limits our coherence time to less than \val{100}{ms}, which is sufficiently high for our purposes. This vapor pressure represents a nearly optimal trade-off between coherence time and MOT loading rate, and occurs conveniently near the room temperature vapor pressure of Rb.

\begin{figure}[t]
\centering{
	\includegraphics[width=1\columnwidth]{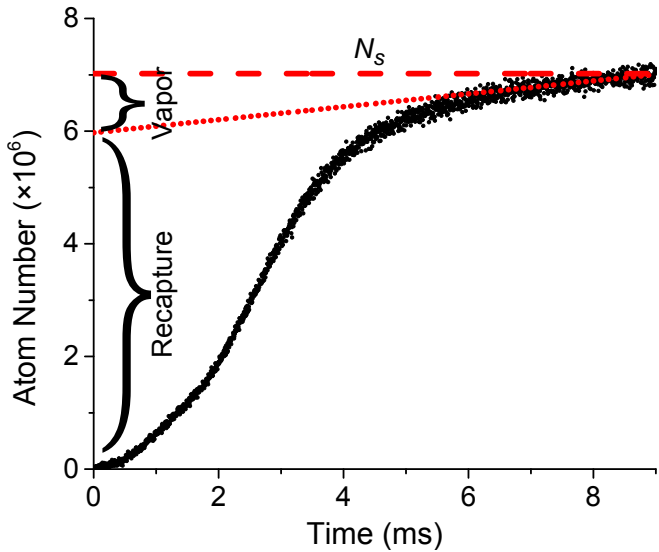}
	\caption{\label{fig:recapture_apd} (Color online) Recapture process during steady-state ensemble exchange. Black dots show the atom number calculated from an avalanche photodiode (APD) signal as atoms are being recaptured and loaded into the trap. The red dashed line indicates the total steady-state atom number $N_s$, while the red dotted line is the vapor loading rate contribution. Many atoms are loaded via recapture over a few milliseconds, followed by vapor loading to replenish the lost atoms. By subtracting out the vapor loading contribution, we determine the net recapture efficiency to be $r = 85\%$. This includes losses due to background collisions during the cycle.}
}
\end{figure}

The loading of two MOTs in close proximity is facilitated by small (approximately \val{16}{mm} length) quadrupole coils located \val{36}{mm} apart, and a custom-designed opto-mechanical frame to support the necessary optics. The sensor head encloses a volume of approximately \val{5000}{cm^3}, excluding components of the high vacuum system. Due to the close proximity of the MOTs, the magnetic field geometry changes based on the relative alignment of the quadrupole field polarities. Aligning the polarities in the same direction causes the magnetic field gradient to vanish at points between the two traps, roughly one quarter the distance between them. We align the polarities in a complementary, opposing geometry to extend the high radial field gradient of \val{9}{G/cm} by a factor of $1.5$. Furthermore, the opto-mechanical frame is machined out of the insulator G10 to diminish eddy currents and allow for fast magnetic field switching, while maintaining a coefficient of thermal expansion similar to that of aluminum. The time-constants for magnetic field switching are \val{14}{\mu s} and \val{37}{\mu s} for bias and quadrupole coils, respectively. This ensures a sufficiently stable magnetic field during the interrogation time of our high bandwidth experiment.

The apparatus uses optical access along four independent geometric planes to densely pack the required 19 optical beams and achieve a compact, modular design. We find that a useful and compact simplification is to employ uncollimated cooling beams that diverge freely from the fiber at a half-angle of 5 degrees \cite{McGuirk}. For each MOT, we use \val{20}{mW} of cooling light locked \val{12}{MHz} red of the $\ket{F=2}\rightarrow\ket{F'=3}$ cycling transition, corresponding to a total saturation parameter $s_0 = 15$. Additionally, \val{1}{mW} of repump light resonant with $\ket{F=1}\rightarrow\ket{F'=2}$ prevents accumulation of population in the dark state.

\autoref{fig:flight} illustrates the launch and recapture process. Prior to the experiment, the quadrupole field is extinguished, and the inner and outer cooling beams are selectively detuned \val{2.2}{MHz} red and blue of the cooling frequency, respectively. This results in optical molasses in a moving frame at \val{2.5}{m/s} towards the center of the cell for both ensembles. The interferometer takes place as the atoms travel ballistically to the other side of the cell over \val{14}{ms}. To facilitate loading for the next shot, at the end of the experiment the traps are turned on and atoms are recaptured in the opposite trap.  The characteristic capture time is given by $\tau_{\mathrm{MOT}} = 2 \hbar k/(\mu' A)$, where $A$ is the magnetic-field gradient and $\mu'$ is the effective magnetic moment of the transition used \cite{Metcalf}. With our typical radial field gradient of \val{9}{G/cm}, we have $\tau_{\mathrm{MOT}} = \val{2.0}{ms}$. Following recapture, atoms are launched again for the next shot of the experiment.

\section{Recapture}

We extend the well-known MOT loading equation \cite{Foot} by including recapture between discrete shots of the experiment. We model the atom number for each shot $n$ as a constant recapture fraction $r_0$, with a loss rate $\beta$ and linear loading rate $\alpha$. Furthermore, we assume a cycle time $T_c$, and a fraction of the cycle $\eta$ reserved for recapture. The atom number may then be modeled as the sequential sum of atoms loaded from vapor, atoms recaptured, and atoms lost to background collisions. If we assume no density-induced losses, the atom number is given by the geometric sequence
\begin{equation}
	N_{n+1} = \alpha \eta T_c + (r_0 - \beta T_c) N_{n},
	\label{eq:discrete_load_geo}
\end{equation}
with solution,
\begin{equation}
	N_n = \frac{\alpha \eta T_c}{1 - r}\left(1 - r^n\right), 
	\label{eq:discrete_load}
\end{equation}
where $r= r_0 - \beta T_c$ is the net recapture efficiency. This assumes that the recapture time $\eta T_c$ is sufficiently larger than the MOT relaxation time, in order to recapture most of the available atoms. When the number of atoms which were not recaptured equals the number of atoms loaded from vapor, we achieve a steady-state atom number $N_s$, given by
\begin{equation}
	N_s = \lim_{n \to \infty} N_n = \frac{\alpha \eta T_c}{1 - r} = \frac{\alpha \eta T_c}{\beta T_c + (1 - r_0)}.
	\label{eq:steady_state}
\end{equation}
For $r_0 < 1$ under low vapor pressure ($\beta T_c \ll 1 - r_0$), the loss rate is dominated by imperfect recapture. Thus, the steady-state atom number $N_s = \alpha\eta T_c / (1 - r_0)$ grows linearly in $\alpha$. Under higher vapor pressure ($\beta T_c \gg 1 - r_0$), the loss rate is dominated by background vapor collisions, resulting in a steady-state atom number $N_s = \eta \alpha / \beta$. Assuming constant trap parameters, $\alpha$ and $\beta$ are both proportional to vapor pressure. Therefore, similar to canonical MOT loading, the ratio $\alpha / \beta$ and thus the total atom number are constant. Note that with $r_0=1, \eta=1$, and in the limit of small $T_c$, $N_n$ becomes the usual MOT loading equation \cite{Foot},
\begin{equation}
	N(t) = \frac{\alpha}{\beta}\left(1 - e^{-\beta t}\right).
	\label{eq:cont_load}
\end{equation}

We test the efficacy of this model with two experiments. In the first, we explore the recapture dynamics of our apparatus, for a \val{20}{ms} cycle time. This is revealed by monitoring the atom number in the trap after the start of recapture until just prior to launch. This experiment is done in steady-state, where the number of atoms lost in recapture is replenished by loading from vapor. \autoref{fig:recapture_apd} demonstrates a large increase in atom number during the recapture process over a millisecond timescale. The large initial increase represents ensemble recapture from the previous shot of the experiment, with $r N_s$ atoms loaded. To achieve steady-state, an additional $\alpha \eta T_c$ atoms are loaded from vapor. By subtracting out the linear contribution and dividing by $N_s$, we calculate a recapture efficiency of $r = 85\%$. From the atom number $N_s = 7 \times 10^6$ and the size of the ensemble from \autoref{fig:flight}, we calculate the peak atom density to be \val{8 \times 10^{9}}{atoms/cm^3}. This is well below the limit for density-induces losses \cite{Foot}, validating the assumption made \autoref{eq:discrete_load_geo}. 

\begin{figure}[t]
\centering{
	\includegraphics[width=1\columnwidth]{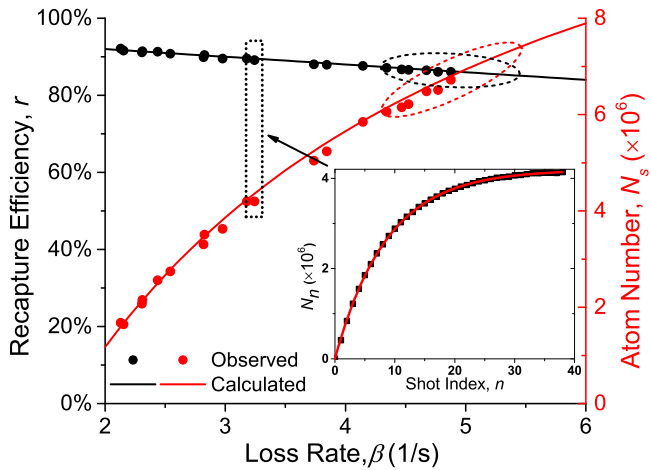}
	\caption{\label{fig:recapture_vapor} (Color online) Inset: Loading of the exchange MOT after starting with the traps off, at $\beta = \val{3.3}{1/s}$. Each data point represents a single shot of the experiment. Fitting the data to \autoref{eq:discrete_load} gives values for $r$ and $N_s$ at any given vapor pressure. Figure: Recapture efficiency $r$ and steady-state atom number $N_s$ as a function of loss rate $\beta$. Circles represent observed data during ensemble-exchange. Solid lines are calculated from MOT loading parameters $\alpha$ and $\beta$, using the base recapture efficiency $r_0$ as a free parameter. The data show that sacrificing a moderate amount of recapture efficiency optimizes the atom number in the cycle, due to a higher loading rate from vapor while the trap is active. The dashed circles show typical operating regimes, as well as the parameters for \autoref{fig:recapture_apd}}.
}
\end{figure}

\begin{figure*}
    \begin{tabular}{lr}
			\includegraphics[width=1\columnwidth]{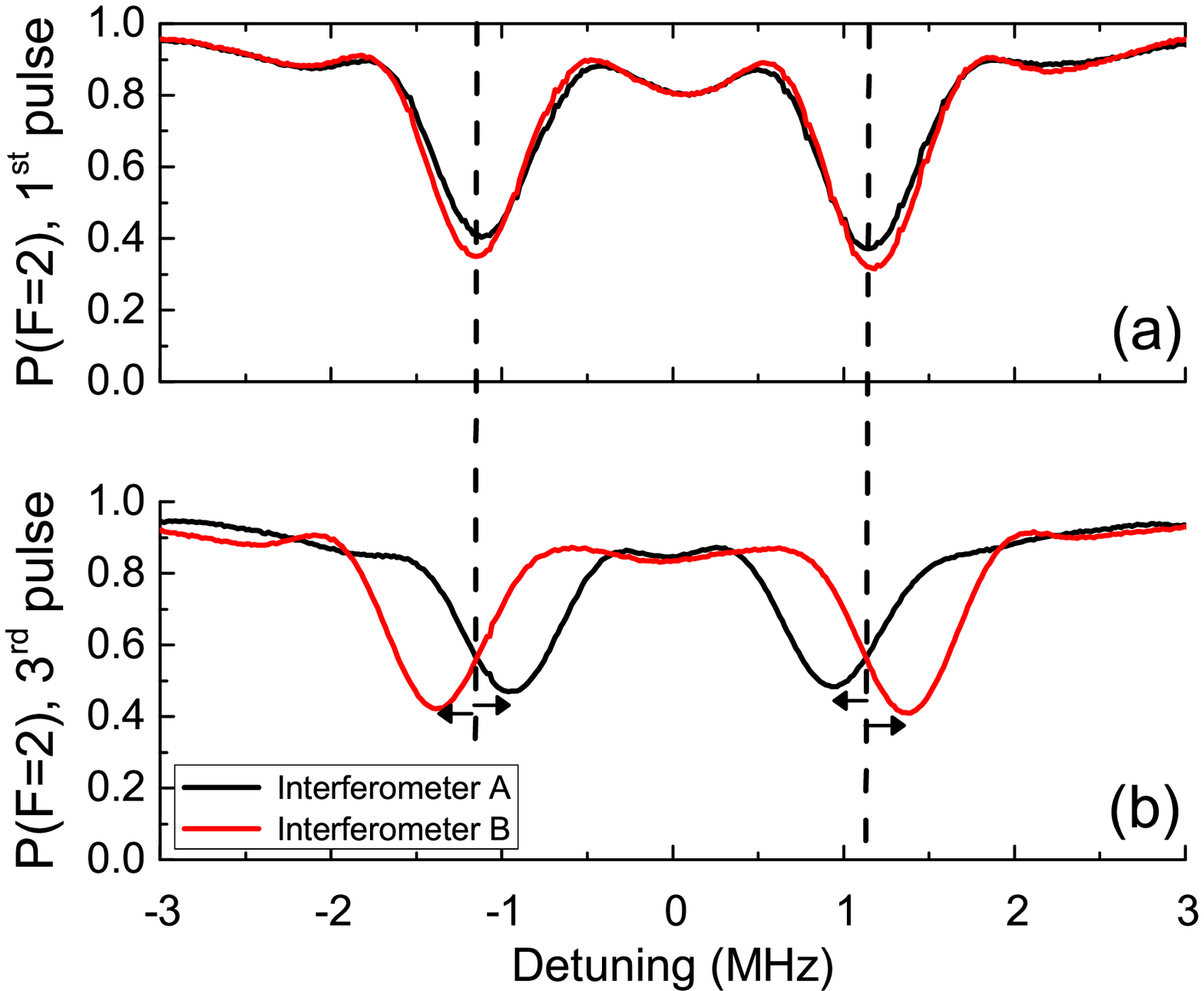} &
			\includegraphics[width=1\columnwidth]{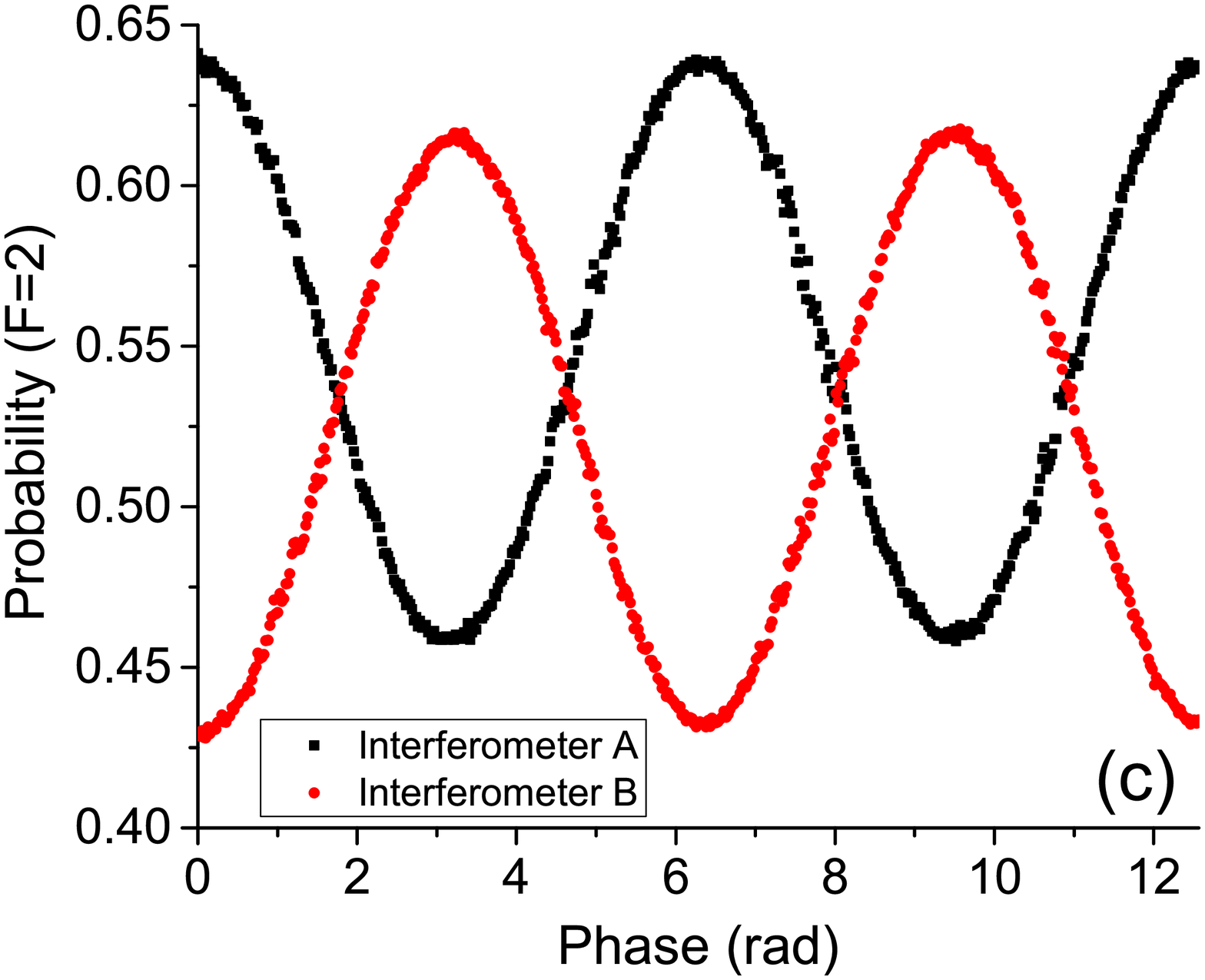} \\			
		\end{tabular}
	  \caption{\label{fig:fringe} (Color online) (a) Transition probability as a function of the two-photon Raman detuning with respect to the hyperfine transition. The $10\degree$ projection of the Raman beam axis along the launch trajectory results in a Doppler shift, causing the resonance conditions of the two $\V{k_e}$ vectors to be separated by \val{2.2}{MHz}. For a given choice of resonance, one interferometer is operated at $+\V{k_e}$ while the other is at $-\V{k_e}$. (b) Gravitational acceleration causes the resonance conditions to shift oppositely for each interferometer, resulting in the second and third pulses being driven off resonance. (c) Sample fringes from the interferometer, generated by scanning the electro-optical phase of the third pulse. The interferometer uses an interrogation time of \val{T = 4.1}{ms}, and data is taken at 60 shots/second.}
\end{figure*}

As a second experiment, we measure the recapture efficiency by fitting \autoref{eq:discrete_load} to the total atom number $N_n$ over multiple shots of the experiment, while the process approaches steady-state. Beginning with zero atoms, ensemble exchange will build the population to a steady-state over 40 cycles. This is depicted in the inset of \autoref{fig:recapture_vapor}. Occasionally triggering the trapping coils on and off allows us to monitor the loading of atoms until steady state. Here, recapture efficiency is revealed by the number of cycles required to reach steady-state.

The recapture efficiency and steady-state atom number are strongly affected by vapor pressure. \autoref{fig:recapture_vapor} depicts the observed and calculated recapture efficiency and steady-state atom number over a range of vapor pressures, characterized by atom loss rate, $\beta$. We demonstrate a recapture efficiency ranging from $85\%$ to $92\%$. The recapture efficiency decreases at higher vapor pressure due to an increased rate of collisions with background atoms during the ensemble's trajectory. However, the increased vapor loading rate results in a net increase of steady-state atom number $N_s$, until an optimum of approximately $85\%$ recapture efficiency is achieved at $\beta = \val{5}{1/s}$. From this fit, we extrapolate the expected recapture efficiency at zero vapor pressure (no background losses) to $96\%$. 

By minimizing the recapture time, we can increase the bandwidth of our experiment and achieve a higher sensitivity. One way to accomplish this is to decrease the trap time-constant $\tau_{\mathrm{MOT}}$ by employing a large magnetic field gradient. However, this limits the maximum atom number in the trap. Accordingly, we employ a dynamic gradient of the quadrupole field. At the start of the recapture, a high field gradient lowers the time constant of the trap allowing for atoms to quickly relax into the trap. The field gradient is slowly ramped down over the remaining recapture time to increase the trap volume as atoms are loaded from vapor. For typical operating parameters, this results in an increase of $10\%$ in steady-state atom number.

\section{Interferometer}

We create the atom interferometer using stimulated Raman transitions. The Raman transitions arise from resonant pulses of light applied at three equispaced times during the ensemble's ballistic trajectory. The Raman beams are seeded from an external cavity diode laser locked \val{900}{MHz} red of the $\ket{F=2}\rightarrow\ket{F'=3}$ transition in $\Rb$. They are generated from the phase coherent zero and first-order sidebands of a fiber-based electro-optic modulator operated at the hyperfine frequency. Following the modulator, the beams are amplified and shuttered by an acousto-optic modulator. Both sidebands are delivered to the apparatus via common optical paths to avoid unwanted relative phase noise. At the apparatus, we separate this light into three beams that are retro-reflected from a shared 2 inch diameter mirror that defines a common inertial reference. The positions of these three beams coincide with the positions of the ensembles at each of the three pulses in the $\pi/2 - \pi - \pi/2$ interferometer pulse sequence. The polarization configuration is lin $\perp$ lin to suppress the Doppler-free transition.

The beams are delivered to the atoms at an angle of $10 \degree$ from the orthogonal of the launch trajectory and are then retro-reflected. This projects a component of the Raman beams along the velocity of the atoms, and the resulting Doppler shift of \val{1.1}{MHz} allows us to resolve both directions of the Raman wavevector $\V{k_e}$ \cite{Stockton2011}. Selecting one of these wavevectors, we drive Doppler-sensitive Raman transitions between the hyperfine levels at three points along the trajectory of the atoms. Due to the asymmetry of the velocity vector of the ensembles, a single detuning results in one interferometer with a vector of $+\V{k_e}$ and another with $-\V{k_e}$. Thus, the acceleration and rotation phase shifts are given by $\phi_+ = \V{k_e}\cdot\V{a} T^2 = (\phi_a + \phi_b)/2$ and $\phi_- = \V{k_e}\cdot(2\V{v} \times \V{\Omega})T^2 = (\phi_a - \phi_b)/2$, respectively.

In our apparatus, gravity is oriented \val{10\degree} from the Raman beam axis, causing a chirp of the Raman transition resonance at a rate of \val{24.7}{kHz/ms}. However, due to the velocity vector asymmetry, this chirp is opposite for the two interferometers. This causes the resonances to separate in frequency space for the second and third pulses. Choosing an intermediate frequency allows both transitions to be resonant, although at a $20\%$ lower efficiency for the final pulse. Raman beam arrangements that create identical $\V{k_e}$ for both interferometers would eliminate this effect for a stationary device. However, a rotation rate of \val{2}{rad/s} in this latter case would cause the same effect.

A typical experiment cycle takes place in \val{16.66}{ms}, corresponding to a data rate of \val{60}{Hz}. Atoms are launched to \val{2.5}{m/s} and cooled over \val{1.5}{ms} to \val{35}{\mu K}. In addition to the repump beam, a $\pi$-polarized, D1, optical pumping beam resonant with the $\ket{F=2}\rightarrow\ket{F'=3}$ transition transfers $92\%$ of the atoms into the magnetically insensitive $\ket{F=2, m_F=0}$ state. A \val{1.5}{G} magnetic field parallel to the Raman beams defines a quantization axis. The atoms are then interrogated by the Raman beams with $T = \val{4.1}{ms}$ at three points along their trajectory. The $\pi$-pulse length is \val{1.6}{\mu s}, allowing for the beams to address a wide velocity class of the atoms for each interferometer.

The population in $\ket{F=2}$ and the total atom number are measured by fluorescence detection. The detection region is smaller than the ensemble size as a means of velocity selection. Thus, only $30\%$ of the total atoms in the ensemble exchange contribute to the interferometer signal, while the remaining $70\%$ facilitate loading for the next shot. We collect $0.5\%$ of the atoms' photon scatter from two \val{100}{\mu s} pulses. The first probe pulse measures the population in $\ket{F=2}$. This is followed by both probe and repumping to measure the total population. The light is coupled into a \val{1}{mm} core multi-mode fiber that delivers it to an APD (Hamamatsu C5460-01). The collimators are aligned off-axis by $2\degree$ to minimize cross talk between the signals. From the ratio of the measured signals, we determine the probability to be in the excited state for each interferometer, from which the phase shifts $\phi_a$ and $\phi_b$ may be calculated. Given a fixed interrogation time, the sensitivity (per $\mathrm{\sqrt{Hz}}$) is optimized with a \val{60}{Hz} ($T_c = \val{16.66}{ms}$) cycle. This restricts the recapture time to \val{6}{ms}, and reduces the steady-state atom number to $4 \times 10^6$. Loading this many atoms in our apparatus from purely vapor would take \val{40}{ms}.

\autoref{fig:fringe} shows a sample fringe from the interferometer, with data taken at \val{60}{Hz}. We measure the phase noise of each interferometer to be $\delta\phi_a = \delta\phi_b = \val{22}{mrad/shot}$. The corresponding acceleration and rotation components of the phase noise are $\delta\phi_+ = \val{19}{mrad/shot}$ and $\delta\phi_- = \val{11}{mrad/shot}$, respectively. The resulting acceleration and rotation sensitivities are \val{0.92}{\ugrtHz} and \val{1.07}{\uradsrtHz} at \val{60}{Hz}. The rotation measurement benefits greatly from common mode noise rejection of the vibrations of the retro-reflecting mirror, with a phase noise approximately $2$ times lower than the acceleration noise. It is worth noting the potential performance of this approach under optimal conditions. If we assume shot-noise limited signal with \val{10^7}{atoms} and 100\% fringe contrast, with identical interrogation time and bandwidth, we calculate optimal sensitivities of \val{21}{\ngrtHz} and \val{44}{\nradsrtHz}. Further optimization of timing parameters is likely to improve sensitivity even further.

The noise in our system derives from reduced fringe contrast and detection noise. The size of the ensemble relative to the size of the Raman beams causes Rabi frequency inhomogeneities, which reduce the overall population transfer efficiency. Additionally, the width of the Raman transition in frequency space must be large enough to address the Doppler shifts from the thermal velocity distribution of the atoms. These effects result in a $\pi$-pulse efficiency of $55\%$ for the first pulse. It is likely that a combination of an expanding position distribution and the \val{24.7}{kHz/ms} gravity-induced Doppler shift reduces the $\pi$-pulse efficiency of the third pulse to $35\%$, resulting in a fringe contrast of $20\%$. Accounting for the fraction of the ensemble imaged and the fringe contrast, $1 \times 10^6$ atoms contribute to the measured noise while $2 \times 10^5$ atoms contribute to the signal. From this, we expect a non-inertial phase noise of \val{5}{mrad} per interferometer, corresponding to $\delta\phi_{\pm} = \val{3.5}{mrad}$. Electronic noise, background scatter, and photon shot noise are negligible. Thus, we attribute the noise in our system to stability of the inertial reference, and frequency and intensity noise of the probe.

\begin{figure}[t]
\centering{
	\includegraphics[width=1\columnwidth]{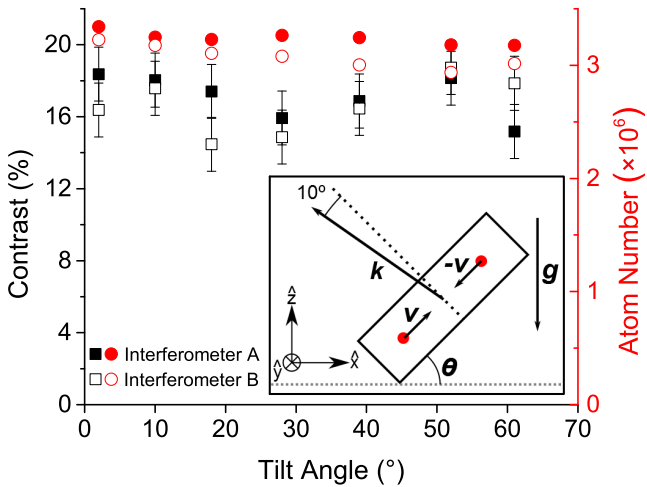}
	\caption{\label{fig:tilt_test} (Color online) Immunity of the atom interferometer to a range of tilt. The tilt direction corresponds to $\theta_y$, thus changing the projection of gravity on the Raman beam axis, as well as the launch axis. Error bars represent the long-term drift seen in the apparatus over \val{1}{hr} when the interferometer is flat on the table. The fringe contrast and steady-state atom number are largely unaffected by the orientation.}
}
\end{figure} 

The most direct route towards enhancing the sensitivity is by improving fringe contrast. It is possible to implement independent detunings for the third pulse by adding a second, independently controllable Raman laser. Additionally, the increased Raman power enables one to reduce Rabi frequency inhomogeneity by increasing Raman beam waist, further improving pulse efficiency. Pulse efficiency can also be improved by using composite pulses \cite{McGuirk, Butts2013, Dunning13}.

\section{Dynamics}

Operating the interferometer in a dynamic environment beyond benign laboratory conditions of \val{1}{\g} and \val{100}{\mu rad/s} demands robustness. Additionally, the apparatus must be able to handle large changes in acceleration (jerk) and rotation (angular acceleration). We motivate further research in this vein by exploring the more dynamic features of our approach compared to conventional atom interferometers.

Gravitational acceleration in conventional atom interferometers with $T \approx \val{100}{ms}$ causes an atomic ensemble to traverse tens of centimeters. Thus, they are designed accounting for a fixed orientation of gravity. Under an arbitrary orientation of gravity, the apparatus will not function due to the ensemble's deflection. A short interrogation time minimizes the deflection and allows for our interferometer to operate in any orientation. To demonstrate this, we rotate the apparatus over a range of angles in the plane defined by $\V{v}$ and $\V{k_e}$. This changes the projection of gravity on these vectors. As shown in \autoref{fig:tilt_test}, the fringe contrast and the atom number remain unchanged to within the measurement uncertainty over a range of angles. The apparatus is designed to exhibit a certain detection and recapture volume. Under a reversal of gravity orientation, the ensemble does not violate these boundaries.

\begin{figure}[t]
\centering{
	\includegraphics[width=1\columnwidth]{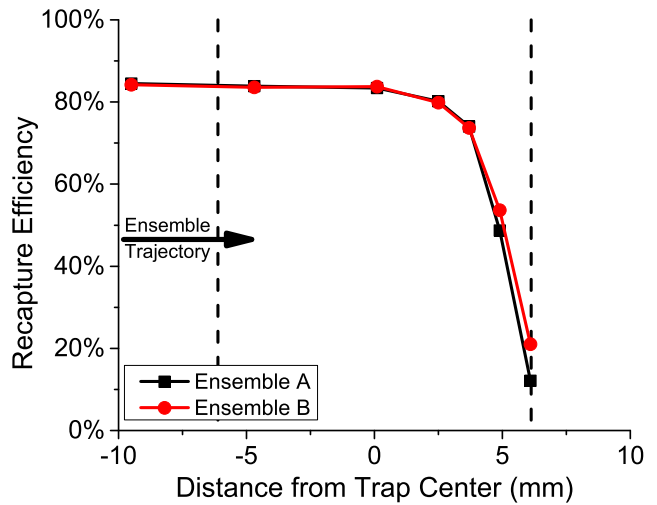}
	\caption{\label{fig:recapture_distance} (Color online) Recapture efficiency over a range of ensemble positions relative to the trap center, at the onset of ballistic recapture. Dashed lines represent the $1/e^2$ diameter of the trapping volume. For early onset (negative position), the ensemble's momentum still carries it into the trap. Recapture is still efficient over a large fraction of the trap volume.}
}
\end{figure}

To estimate the acceleration and rotation rate the ensemble exchange can endure, we measure the recapture efficiency as a function of ensemble distance from the trap center at the onset of recapture. By triggering the trap at different delay times following the launch, we simulate recapture at different locations along the launch vector, $\V{\hat{x}}$ (\autoref{fig:recapture_distance}). While the launch velocity direction causes an asymmetry where early activation still allows the atoms to relax into the trap, results for a delayed activation should mirror deflections along the $\V{\hat{y}}$ and $\V{\hat{z}}$ axes. We empirically measure the recapture radius along $\unitV{x}$ to be \val{5}{mm}, where the recapture efficiency falls below $50\%$. Dynamic conditions as large as \val{10}{\g} of acceleration and \val{20}{rad/s} of rotation would be necessary to replicate this level of excursion. 

Although the ensemble exchange tolerates a large range of inertial conditions, successfully tracking the inertial signals requires that jerk and angular acceleration be limited to an amount equivalent to a $\pi/2$ interferometer phase shift over one interferometer cycle. This corresponds to a jerk of $\magV{j} = \pi/(2 \magV{k_e} T^2 T_c)$, and an angular acceleration of $\magV{\alpha} = \pi/(4 \magV{k_e} \magV{v} T^2 T_c)$. Exceeding these rates can be enabled by using complementary auxiliary sensors.

\begin{figure}[t]
\centering{
	\includegraphics[width=1\columnwidth]{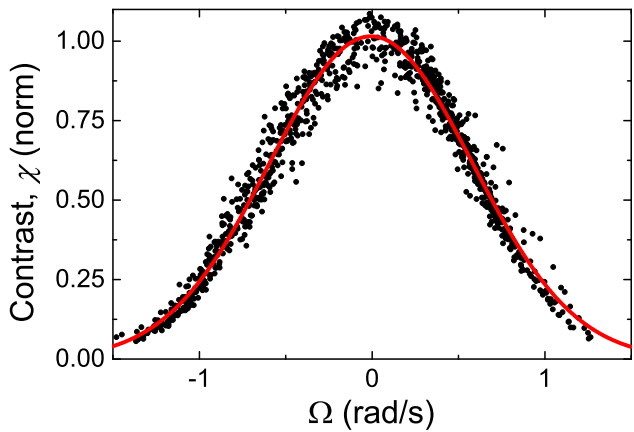}
	\caption{\label{fig:contrast_loss} (Color online) Contrast loss under high rotation rate. The black points are the observed fringe contrast as the apparatus was rotated in the $\pm\unitV{g}$ direction, up to \val{\pm1.5}{rad/s}. The red line is a Gaussian fit, with a width of $\sigma = \val{0.60(1)}{rad/s}$.}
}
\end{figure}

An alternative rotation rate limit arises from phase gradients in the ensemble due to the thermal velocity distribution, leading to loss of contrast. Each atom derives an independent rotational phase shift due to its thermal velocity coupled with the rotation rate. For a given characteristic velocity width $\sigma_v$, when the rotation rate exceeds the condition $\magV{\Omega} > \pi/(2 \magV{k_e} \sigma_v T^2)$, large inhomogeneous dephasing occurs \cite{Gustavson00}. Given a rotation of magnitude $\Omega$ at an angle of $\theta$ from $\mathbf{\hat{k}_e}$, and an ensemble temperature of $\mathcal{T}$, the decay of the fringe contrast $\chi$ is given by,
\begin{eqnarray}
  \chi(\Omega) = \exp\left(\frac{-\Omega^2}{2\sigma^2}\right), \nonumber\\
	\sigma = \frac{1}{2}\sqrt{\frac{m}{k_{\mathrm{B}} \mathcal{T}}}\frac{1}{\magV{k_e} T^2 \sin{(\theta)}},
	\label{eq:rot_gauss}
\end{eqnarray}
where $m$ is the mass of $\Rb$, and $k_{\mathrm{B}}$ is the Boltzmann constant. 

We demonstrate this effect by placing the apparatus on a \val{22}{cm} diameter rotation platform located outside of the magnetic shielding. The experiment is configured to rotate about $\pm\unitV{g}$ in \autoref{fig:apparatus}, or $\theta = 10 \degree$. The angular position of the platform is recorded via a rotary encoder. \autoref{fig:contrast_loss} shows measurements of the fringe contrast while rotating the apparatus in the $\pm\unitV{g}$ direction up to \val{\pm1.5}{rad/s}. We observe a minimal change in atom number of a couple percent at \val{1}{rad/s} for a duration of \val{1}{s}. At sustained rates where the ensemble is allowed to hit steady-state, we expect the atom number to be slightly less. As expected, the contrast loss is Gaussian with respect to the rotation rate $\Omega$. The characteristic width of the distribution is $\sigma = \val{0.60(1)}{rad/s}$, which corresponds to a temperature of \val{36}{\mu K}. This is in agreement with independent temperature measurements. The most severe limit would arise if we rotated the apparatus perpendicular to $\mathbf{\hat{k}_e}$, which would cause the contrast to vanish $\sin{(90\degree)}/\sin{(10\degree)} \approx 6$ times faster. This limit can be extended by cooling to lower temperature or by placing the apparatus on gimbals. Although the ensemble contrast is diminished, the individual atoms still maintain phase coherence and inertial purity. If the ensemble meets the condition $2 T \sigma_v \gg \sigma_{x_0}$, where $\sigma_{x_0}$ is the initial characteristic position width, the position of each atom during detection corresponds strongly with its velocity. Then, the inertial information for each velocity class may be retrieved with a sufficiently well-resolved detector \cite{Sugarbaker2013}. 

\section{Conclusion}

We have demonstrated a high data-rate atom interferometer capable of simultaneous acceleration and rotation measurements at \val{60}{Hz}, with sensitivities suitable for a wide application space. The system makes use of the ensemble exchange technique to launch and recapture atoms at a high data-rate, orders of magnitude higher than typical LPAI systems. We modeled the atom number during ensemble exchange assuming a fixed recapture fraction, and validated the model with experiments. Additionally, we characterized our interferometer and estimated the potential performance of our design. We showed immunity of recapture efficiency and interferometer contrast over a range of acceleration and rotation values.

Our ensemble exchange method enables a path towards a compact atomic inertial measurement unit. Further work is needed to miniaturize the remainder of the system and achieve the ultimate sensitivity. We consider a device with sensitivities of \val{{\sim}10}{\ngrtHz} and \val{{\sim}10}{\nradsrtHz}, and robust up to \val{{\sim}10}{\g} and \val{{\sim}10}{rad/s} to be feasible in an engineered package that is significantly smaller than our apparatus. Through careful study of the apparatus' dynamics, we may model the sensor response in various scenarios and enable a fielded device.
 
\begin{acknowledgments}
We gratefully thank G. Burns for experimental support and Y.-Y. Jau for helpful discussions. We would also like to thank D. Butts, D. Johnson, and R. Stoner at Draper Laboratory for helpful discussions regarding development of an inertial measurement unit using this technique. This work was supported by the Laboratory Directed Research and Development (LDRD) program at Sandia National Laboratories. Sandia National Laboratories is a multi-program laboratory managed and operated by Sandia Corporation, a wholly owned subsidiary of Lockheed Martin Corporation, for the U.S. Department of Energy's National Nuclear Security Administration under Contract No. DE-AC04-94AL85000.
\end{acknowledgments}

\bibliographystyle{apsrev4-1}
\bibliography{sandia_gyro}

\end{document}